\documentclass[manuscript]{aastex61}

\newcommand\aastex{AAS\TeX}

\newcommand{\fpp}[2]{\frac{\partial #1}{\partial #2}}
\newcommand{\vctr}[1]{\mbox{\boldmath $#1$}}
\newcommand{\mrm}[1]{\mathrm{#1}}

\shorttitle{\aastex\ RC-RTI}
\shortauthors{Kaneko and Yokoyama}

\begin{document}

\title{Impact of Dynamic State on the Mass Condensation Rate of Solar Prominences}

\correspondingauthor{Takafumi Kaneko}
\email{kaneko@isee.nagoya-u.ac.jp}

\author[0000-0002-7800-9262]{Takafumi Kaneko}
\affil{Institute for Space-Earth Environmental Research, Nagoya University,
  Furo-cho, Chikusa-ku, Nagoya, Aichi 464-8601, Japan}

\author{Takaaki Yokoyama}
\affil{Department of Earth and Planetary Science,The University of Tokyo,
7-3-1 Hongo, Bunkyo-ku, Tokyo, 113-0033, Japan}

\begin{abstract}
  The interiors of quiescent prominences are filled with turbulent flows.
  The evolution of upflow plumes, descending pillars, and vortex motions
  has been clearly detected
  in high-resolution observations. The Rayleigh--Taylor instability
  is thought to be a driver of such internal flows.
  Descending pillars are related to the mass budgets of prominences.
  There is a hypothesis of dynamic equilibrium
  where the mass drainage via descending pillars and
  the mass supply via radiative condensation are balanced
  to maintain the prominence mass; however, the background physics connecting
  the two different processes is poorly understood.
  In this study, we reproduced the dynamic interior of a prominence
  via radiative condensation and the mechanism similar to the Rayleigh--Taylor instability
  using a three-dimensional magnetohydrodynamic simulation
  including optically thin radiative cooling and nonlinear anisotropic
  thermal conduction.
  The process to prominence formation in the simulation
  follows the reconnection--condensation model, where topological change
  in the magnetic field caused by reconnection leads to radiative condensation.
  Reconnection is driven by converging motion at the footpoints of the coronal arcade fields.
  In contrast to the previous model, by randomly changing the speed of the footpoint motion
  along a polarity inversion line, the dynamic interior of prominence is successfully reproduced.
  We find that the mass condensation rate of the prominence
  is enhanced in the case with dynamic state.
  Our results support the observational hypothesis that 
  the condensation rate is balanced with the mass drainage rate
  and suggest that a self-induced mass maintenance mechanism exists. 
\end{abstract}


\section{Introduction} \label{sec:intro}

Solar prominences or filaments are cool dense plasma clouds
suspended in coronal magnetic fields.
In limb observations, quiescent prominences appear to be
composed of fine vertical threads
\citep{Engvold1976SoPh,Engvold1981SoPh,Berger2008ApJ,Berger2010ApJ,Chae2008ApJ,Chae2010ApJ}.
Both upflows and downflows are present in prominences
\citep{Zirker1994SoPh,Zirker1998Natur}
and show turbulent characteristics  
\citep{Freed2016ApJ,Hillier2017AA}.
Rising dark plumes and descending pillars associated with internal flows
have been clearly detected via high-resolution observations
of the Solar Optical Telescope on the {\it Hinode} satellite
\citep{Berger2008ApJ,Berger2010ApJ,Berger2011Natur,Berger2017ApJ,Chae2008ApJ,Chae2010ApJ}.

The downflows and upflows are related to the mass budget of the prominences.
Prominences drain their mass via descending pillars \citep{Hillier2012ApJb,Liu2012ApJ}
and obtain mass from upflow plumes \citep{Berger2010ApJ},
chromospheric jets \citep{Chae2003ApJ},
flux emergence \citep{Okamoto2008ApJ,Okamoto2009ApJ,Okamoto2010ApJ},
and radiative condensation in the corona \citep{Berger2012ApJ,Liu2012ApJ}.
Though it is difficult to estimate the mass supply or drainage rate
of each process in most prominences,
the condensation rate and the mass drainage rate have been reported to be comparable
and temporally correlated with each other in one in-situ prominence
formation event \citep{Liu2012ApJ}.
According to estimates,
the time scale of mass cycle (the total mass divided by the mass drainage rate)
is less than one hour,
which is much shorter than the typical lifetime of prominences
(a few hours to months; \citet{Mackay2010SSRv}).
This result suggests a dynamic equilibrium 
where the prominence mass is maintained by the complete balance between
continuous and substantial mass supply and drainage.
To reveal the mass maintenance of prominences,
the physical interactions between the mass drainage and the condensation,
as well as the detailed mechanisms of each process, need to be investigated.

In most previous studies, the mass drainage and the condensation have been separately investigated.
For example, the Rayleigh--Taylor instability is thought to be a mechanism
driving upflow plumes and descending pillars \citep{Berger2010ApJ,Ryutova2010SoPh}.
Previous theoretical studies of the Rayleigh--Taylor instability
using ideal magnetohydrodynamic (MHD) simulations
succeeded in reproducing upflows and downflows with speeds
consistent with observational values 
\citep{Hillier2011ApJ,Hillier2012ApJa,Hillier2012ApJb,KeppensXia2015ApJ,XiaKeppens2016RT};
however, these simulations neglected the mass supply to the prominence
via radiative condensation due to their use of the adiabatic assumption.
  
Radiative condensation has been investigated as a mass supply mechanism for prominences.
Several models with different drivers have been proposed.
One is the evaporation--condensation model
in which chromospheric evaporation injects mass into the coronal loops,
leading to radiative condensation.
Early studies using one-dimensional hydrodynamic simulations,
where the magnetic fields were assumed to be fixed, were not able to discuss interactions with the Rayleigh--Taylor instability \citep{Antiochos1991ApJ,Antiochos1999ApJ,Karpen2001ApJ,Karpen2003ApJ,Karpen2005ApJ,Karpen2006ApJ,KarpenAntiochos2008ApJ,Xia2011ApJ,Luna2012ApJ}.
Recently, this model was investigated using multi-dimensional MHD simulations including radiative cooling and thermal conduction.
In two-dimensional simulations, prominences were reproduced in the form of a stationary slab
without a dynamic interior \citep{XiaChenKeppens2012ApJ,KeppensXia2014ApJ}.
Using three-dimensional simulation, \citet{XiaKeppens2016ApJ}
found that a dynamic fragmented prominence is reproduced with this model.
Moreover, in their simulations, the total mass of the simulated prominence is maintained via the balance between the mass input from chromospheric evaporation and the mass drainage from prominences,
showing a solution achieving dynamic equilibrium.
The mass drainage in their simulation was driven by overdense clusters of fragmented condensations that were not sustained by the weak coronal magnetic fields.
The role of the Rayleigh--Taylor instability in this simulation was not clear.
In addition, even though parameterized heating is an important factor to determine the mass flux
of the evaporated flow in this model,
neither observations nor theories have yet to guarantee its presence.
It is still unclear whether the dynamic equilibrium is self-induced
or contingent on the parameter settings.
  
Another condensation model, the reconnection--condensation model, has
also been proposed in previous theoretical studies
\citep{Pneuman1983SoPh,KanekoYokoyama2015ApJ,KanekoYokoyama2017ApJ}.
This model was recently
demonstrated using three-dimensional MHD simulations including thermal conduction
and optically thin radiative cooling in a study by \citet{KanekoYokoyama2017ApJ}.
In their simulations, a flux rope is formed in the sheared arcade fields by reconnection
via converging motion at the footpoints of the arcade fields.
Radiative cooling is enhanced due to dense coronal plasmas trapped inside the flux rope;
however, thermal conduction along the long coronal loops after reconnection
can not compensate for the radiative losses, leading to radiative condensation.
The simulated prominence ended up in a stationary slab without the dynamic features
seen in observations. The convergence of magnetic patches of opposite polarities
has been detected below prominences in observations of the photospheric magnetic field
\citep{Rondi2007AA,Schmieder2014AA,Yang2016ApJ}.
The issue remaining for this model, therefore, is to reproduce the internal dynamics
associated with the mass budget of prominences.

In this study,
we show that random footpoint motion leads to the formation of prominences
with dynamic fine structures by radiative condensation and the mechanism similar to
the Rayleigh--Taylor instability
in the framework of the reconnection--condensation model.
Moreover, we show that the dynamic state has an impact on the growth of radiative condensation
by comparing simulations with and without the dynamic state.
The numerical settings are described in Section \ref{sec:setting}.
The results of the simulations are shown in Section \ref{sec:result}.
We discuss the results in Section \ref{sec:discussion},
and the conclusions are given in Section \ref{sec:conclusion}.
  
\section{Numerical settings} \label{sec:setting}

The simulation domain is a rectangular box 
whose Cartesian coordinates $(x,y,z)$  
extend to $-12$ $\mathrm{Mm} < x < 12$ $\mathrm{Mm}$, 
$0 < y < 50$ $\mathrm{Mm}$, and $0 < z < 24$ $\mathrm{Mm}$, 
where the $y$-direction corresponds to the height and the $xz$-plane 
is parallel to the horizontal plane.
The initial corona is under hydrostatic stratification
with a uniform temperature ($T_\mathrm{cor}=1~\mathrm{MK}$)
and uniform gravity ($g_\mathrm{cor}=270~\mathrm{m/s^{2}}$).
The initial density profile is given as
\begin{equation}
  n=n_{\mathrm{cor}}\exp \left[ -\frac{y}{L_{s}}\right],
\end{equation}
where $n$ is number density,
$n_{\mathrm{cor}}=10^{9}~\mathrm{cm^{-3}}$ is the number density at the coronal bottom,
and $L_{s}=k_{B}T_{\mathrm{cor}}/(mg_{\mathrm{cor}})=30~\mathrm{Mm}$
is the coronal scale height,
where $k_{B}$ is the Boltzmann constant and $m$ is the mean molecular mass.
The mean molecular mass of a prominence depends on the helium abundance and
the ionization degree. An accurate treatment of the ionization degree
requires non-LTE (local thermodynamic equilibrium) modeling. 
For simplicity, we set $m=m_{p}$, where $m_{p}$ is the proton mass.

The initial magnetic field is a linear force-free arcade given as
\begin{eqnarray}
  B_{x} &=& -\left( \frac{2L_{a}}{\pi a}\right)B_{a} 
  \cos \left( \frac{\pi x}{2L_{a}} \right) \exp \left[ -\frac{y}{a} \right], \\
  B_{y} &=& B_{a}\sin \left( \frac{\pi x}{2L_{a}} \right) \exp \left[ -\frac{y}{a} \right], \\
  B_{z} &=& -\sqrt{1-\left( \frac{2L_{a}}{\pi a}\right)^{2}}  
  B_{a}\cos \left( \frac{\pi x}{2L_{a}} \right) \exp \left[ -\frac{y}{a} \right],
\end{eqnarray}
where $B_{a}=3~\mathrm{G}$, $L_{a}=12~\mathrm{Mm}$, and $a=30~\mathrm{Mm}$.
The polarity inversion line (PIL) on the surface $y=0$ is located along $x=0$.

The three-dimensional MHD equations including nonlinear anisotropic
thermal conduction and optically thin radiative cooling are as follows:
\begin{equation}
  \fpp{\rho }{t}+\nabla \cdot \left(\rho \vctr{v}\right)=0,
  \label{eq_mass_nc}
\end{equation}
\begin{equation}
  \fpp{\left(\rho \vctr{v}\right)}{t}+\nabla \cdot \left( \rho \vctr{v}\vctr{v}
  +p\vctr{I}-\frac{\vctr{B}\vctr{B}}{4\pi }+\frac{B^2}{8\pi
  }\vctr{I} \right) -\rho \vctr{g}=0,
  \label{eq_momentum_nc}
\end{equation}
\begin{eqnarray}
  \frac{\partial }{\partial t}\left( e_\mrm{th}+\frac{1}{2}\rho
  \vctr{v}^{2} +\frac{B^2}{8\pi }
  \right)
  +\nabla \cdot \left[ \left( e_\mrm{th} + p +\frac{1}{2}\rho
    \vctr{v}^{2}\right)\vctr{v}+\frac{c}{4\pi }\vctr{E} \times \vctr{B}
    \right] \nonumber \\
  =\rho \vctr{g}\cdot \vctr{v}
  +\nabla \cdot \left(\kappa T^{5/2}\vctr{b}\vctr{b}\cdot \nabla T \right)
  -n^{2}\Lambda (T)+H,
  \label{eq_energy_nc}
\end{eqnarray}
\begin{equation}
  e_\mrm{th}=\frac{p}{\gamma -1},
  \label{eth}
\end{equation}
\begin{equation}
  T=\frac{m}{k_{B}}\frac{p}{\rho },
  \label{te_def}
\end{equation}
\begin{equation}
  \fpp{\vctr{B} }{t}=-c\nabla \times \vctr{E},
  \label{eq_induction}
\end{equation}
\begin{equation}
  \vctr{E}=-\frac{1}{c}\vctr{v}\times \vctr{B}
  +\frac{4\pi \eta }{c^{2}} \vctr{J},
  \label{eq_ohm}
\end{equation}
\begin{equation}
  \vctr{J}=\frac{c}{4\pi }\nabla \times \vctr{B},
  \label{eq_current}
\end{equation}
where $\mrm{\kappa =2\times 10^{-6}~erg~cm^{-1}~s^{-1}~K^{-7/2}}$
is the coefficient of thermal conduction, $\vctr{b}$
is a unit vector along the magnetic field,
$\Lambda (T)$ is the radiative loss function of an
optically thin plasma, $H$ is the background heating rate per volume,
and $\eta $ is the magnetic diffusion rate.
We use the same radiative loss function as that used
in \citet{KanekoYokoyama2017ApJ}.
The radiative loss function under $10^{4}~\mathrm{K}$ is assumed to have
a dependence of $T^{3}$.
The background coronal heating is taken to be proportional to the magnetic energy density and
is given as
\begin{equation}
  H=\alpha_{H} B^{2},
\end{equation}
where $B$ is the magnetic field strength and
$\alpha _{H}= 4.8 \times 10^{-7}~\mathrm{s^{-1}}$ is a constant coefficient.
The value of $\alpha _{H}$ is computed to satisfy the condition for
thermal equilibrium for a uniform temperature: 
\begin{equation}
  n^{2}\Lambda (T_\mathrm{cor})=H,
\end{equation}
where $a=L_{s}$
(see also Eqs. (13)--(15) in \citet{KanekoYokoyama2017ApJ} for the equilibrium condition).

To drive reconnection,
the footpoint velocities perpendicular and parallel to the PIL are given in $y<0$ as
\begin{eqnarray}
  v_{x} &=& -v_{0}(t)\sin \left( \frac{\pi x}{2L_{a}} \right)f(z),
  \label{nc2_rand_x} \\
  v_{y} &=& 0, \label{nc2_rand_y}\\
  v_{z} &=& v_{x}, \label{nc2_rand_z}
\end{eqnarray}
where $t$ represents time, $v_{0}(t)$ is the speed depending on time,
and $f(z)$ represents a random number with an amplitude of $0.5 \leq f(z) \leq 1.5$.
The velocity component $v_{x}$ represents the converging motion used to drive reconnection.
The component $v_{z}$ represents the anti-shearing motion necessary
to trigger radiative condensation and to prevent the eruption of a flux rope
\citep{KanekoYokoyama2017ApJ}.
The speed $v_{0}(t)$ is given as
\begin{eqnarray}
  v_{0}(t) &=& v_{00}, ~~(0<t<t_{1}) \label{eqs:vt1}\\
  v_{0}(t) &=& v_{00}\frac{t_{2}-t}{t_{2}-t_{1}}, ~~(t_{1} \le t \le t_{2}) \label{eqs:vt2}\\
  v_{0}(t) &=& 0,~~(t \ge t_{2}) \label{eqs:vt3}
\end{eqnarray}
where $v_{00}=12~\mathrm{km/s}$, $t_{1}=1200~\mathrm{s}$, and $t_{2}=1440~\mathrm{s}$.
For three grids below $y=0$, the magnetic fields are numerically calculated
with the given velocities Eqs. (\ref{nc2_rand_x})--(\ref{eqs:vt3})
based on the induction equation and a free boundary condition at the bottom boundary.
Via this manipulation, the given footpoint velocity is smoothly adopted
onto the domain solving the full MHD equations.
The gas pressure and density below $y=0$ are assumed to
be in hydrostatic equilibrium at a uniform temperature of 
$1~\mathrm{MK}$. The free boundary condition is applied to all the variables 
at the top boundary.
The anti-symmetric boundary condition is 
applied to $v_{x}$, $v_{z}$, $B_{x}$, and $B_{z}$, and the symmetric boundary condition is 
applied to the other variables at the boundaries in the $x$-direction.
Periodic boundary conditions are applied to all variables 
at the boundaries in the $z$-direction under the assumption that a flux rope 
sustaining a prominence is sufficiently long.

For fast magnetic reconnection,
we adopt the following form of the anomalous resistivity
\citep[e.g.][]{YokoyamaShibata1994ApJ}:
\begin{eqnarray}
  \eta & = & 0, ~~~\left(J < J_{c}\right)\\
  \eta & = & \eta _{0}\left(J/J_{c}-1\right)^{2}, ~~~\left(J \ge J_{c}\right)
\end{eqnarray}
where $\eta _{0}=\mrm{3.6\times 10^{13}~cm^{2}~s^{-1}}$ and
$J_{c}=\mrm{25~erg^{1/2}~cm^{-3/2}~s^{-1}}$.
We restrict $\eta \leq \eta _{\mrm{max}}=\mrm{1.8\times 10^{14}~cm^{2}~s^{-1}}$.

For comparison, we performed an additional simulation for
footpoint motions without random components by fixing $f(z_{k})=1$.
This simulation corresponds to a 2.5-dimensional simulation in the $xy$-plane
where the translational symmetry is constrained in the $z$-direction (along the PIL).
Hereafter, we call the case with random footpoint motion case R,
and the case with footpoint motion without random components case N.
For case N, the basic parameters and boundary conditions, except for the velocities
at the footpoints, are the same as those for case R.

\section{Results} \label{sec:result}

\begin{figure}[htbp]
  \begin{center}
    \includegraphics[bb=0 0 452 496]{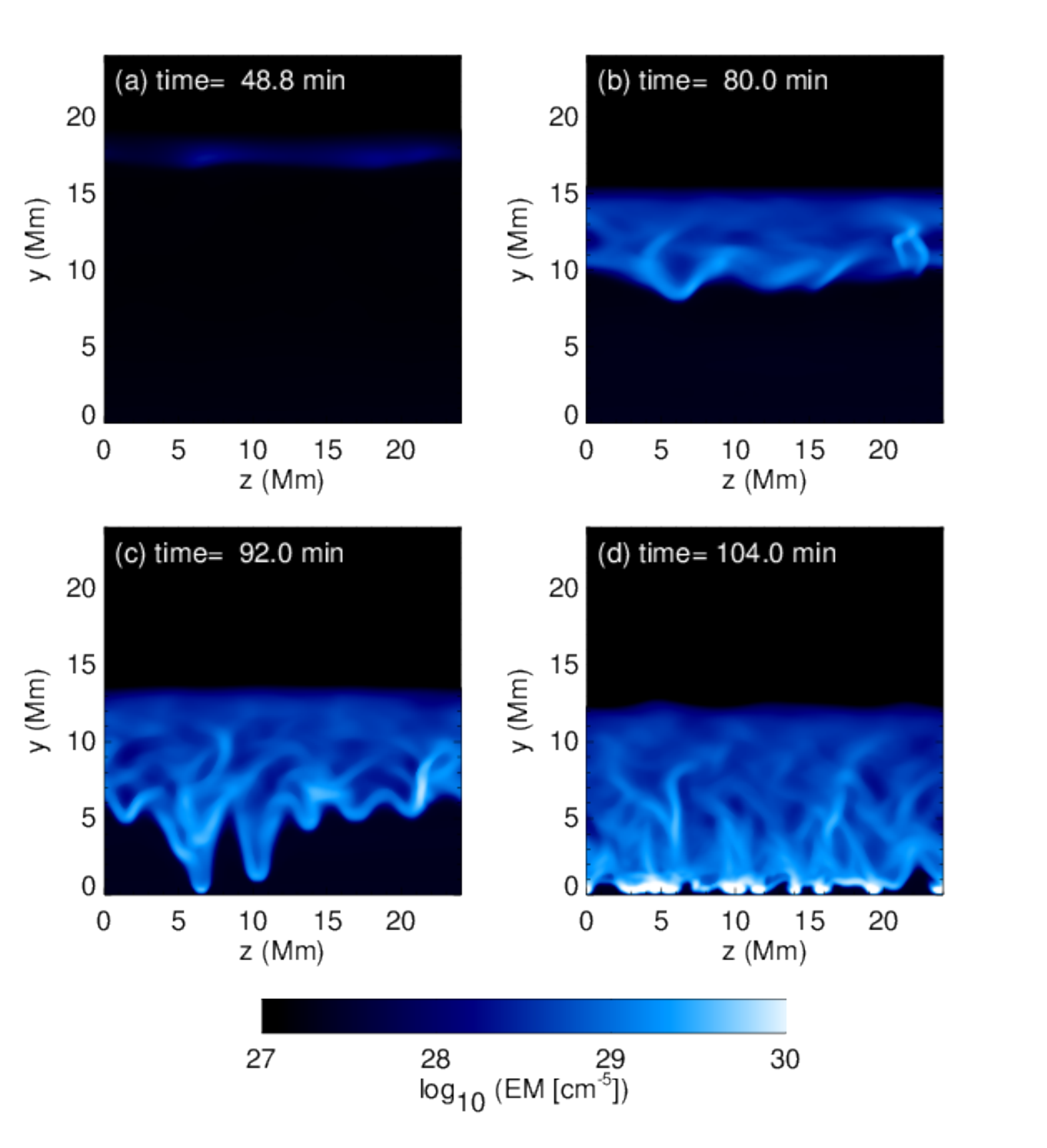}
    \caption{Time evolution of the emission measure along the $x$-axis for case R.}
    \label{fig:ems3d}
  \end{center}
\end{figure}

Figure \ref{fig:ems3d} shows snapshots of the emission measure, which is defined as
\begin{equation}
  \mathrm{EM}(y,z)=\int n(x,y,z)^{2} dx
\end{equation}
when observed with the line of site in the $x$-direction, for case R.
The process leading to radiative condensation is the same
as that in previous studies \citep{KanekoYokoyama2015ApJ,KanekoYokoyama2017ApJ}.
A flux rope is created by the reconnection of the sheared arcade fields 
via converging footpoint motion.
Radiative cooling inside the flux rope is enhanced
because the dense plasmas in the lower corona are trapped.
Due to the topology change of the magnetic fields,
the relaxation effect via thermal conduction becomes ineffective
at compensating for radiative loss inside the flux rope, leading to radiative condensation.
As shown in Fig. \ref{fig:ems3d}(a), the density of the condensation is nonuniform
in the $z$-direction because random components of the footpoint motion add fluctuation to
the interior of the flux rope prior to condensation
(note that the footpoint motion is stopped prior to condensation).
As the mass of the prominence increases, multiple spikes grow 
(Figs. \ref{fig:ems3d}(b) and \ref{fig:ems3d}(c)).
Thin vertical threads are formed after the spikes reach the bottom boundary
(Fig. \ref{fig:ems3d}(d)).

\begin{figure}[htbp]
  \begin{center}
    \includegraphics[bb=0 0 509 254]{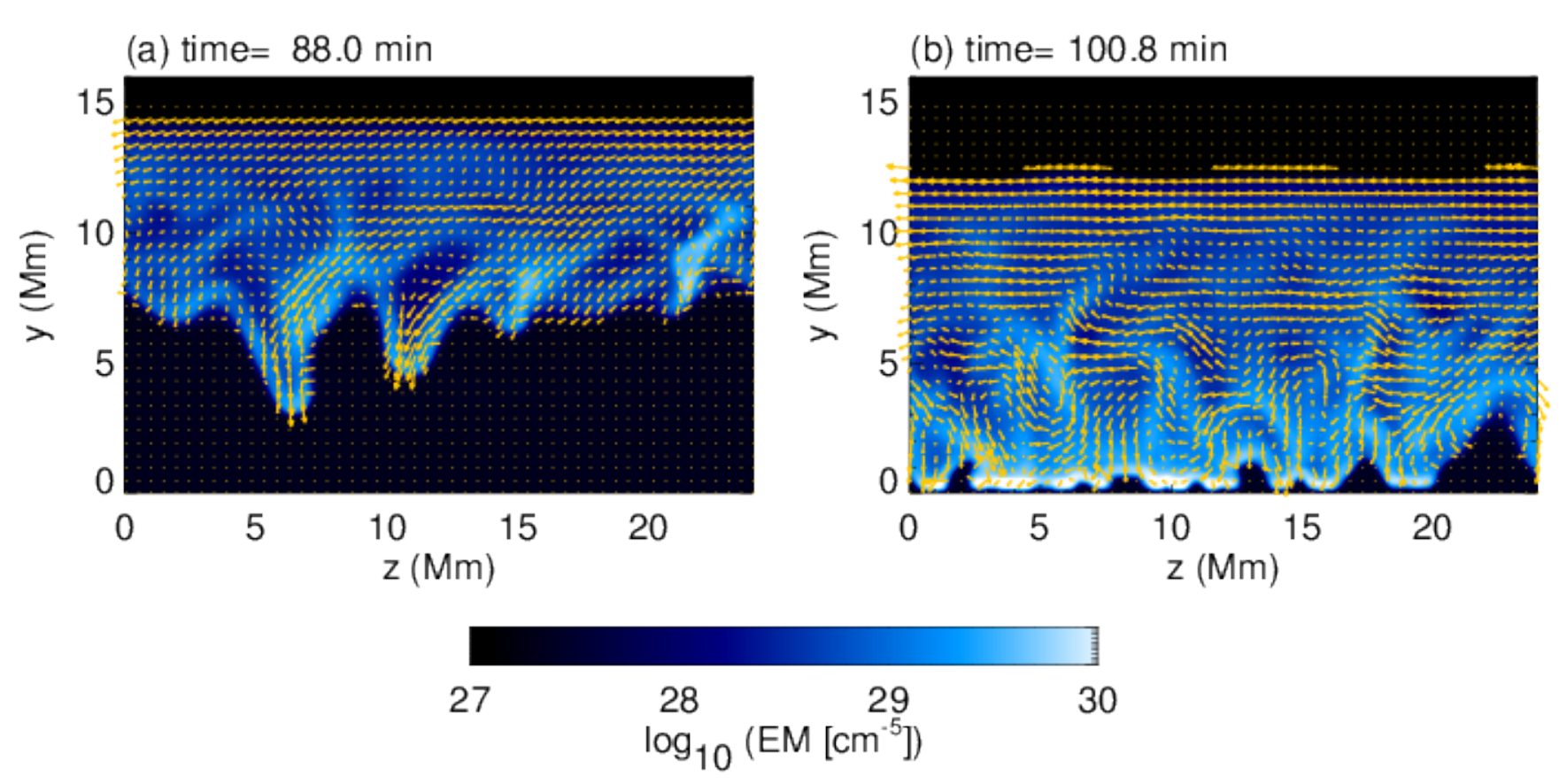}
    \caption{Flows inside the prominence overlaid on the emission measure
      before and after the spikes reach the bottom boundary for case R.
      The arrows represents the velocities $(V_{y}^{p},V_{z}^{p})$
      as defined in Eqs. (\ref{eq:vyp}), (\ref{eq:vzp}) and (\ref{eq:massx}).}
    \label{fig:ems3d_vel}
  \end{center}
\end{figure}

Figure \ref{fig:ems3d_vel} shows the velocity fields inside the prominence, which is defined as
\begin{equation}
  V_{y}^{p}(y,z)=\frac{1}{M_{x}}\int _{T<10^{5}\mathrm{K}} \rho v_{y} dx,
  \label{eq:vyp}
\end{equation}
\begin{equation}
  V_{z}^{p}(y,z)=\frac{1}{M_{x}}\int _{T<10^{5}\mathrm{K}} \rho v_{z} dx,
  \label{eq:vzp}
\end{equation}
\begin{equation}
  M_{x}=\int _{T<10^{5}\mathrm{K}} \rho dx.
  \label{eq:massx}
\end{equation}
The strong downflows are concentrated in the descending spikes
(Fig. \ref{fig:ems3d_vel}).
The downward speed of the spikes is approximately $12~\mathrm{km/s}$,
which is consistent with
observational values of $10$--$15~\mathrm{km/s}$
\citep{Berger2008ApJ,Chae2010ApJ,Hillier2012ApJb}.
The spikes are reflected at the bottom boundary and upflows
or vortex motions are created (see around $(y,z)=(4~\mathrm{Mm},5~\mathrm{Mm})$
in Fig. \ref{fig:ems3d_vel}(b)).
The spikes are squeezed via the interactions of the flows.
Eventually, thin vertical threads form along the paths of the descending spikes.
The widths of the threads in our simulation are approximately $1000~\mathrm{km}$,
comparable to the observed width of $600~\mathrm{km}$ \citep{Chae2010ApJ}.

\begin{figure}
  \begin{center}
    \includegraphics[bb=0 0 315 626]{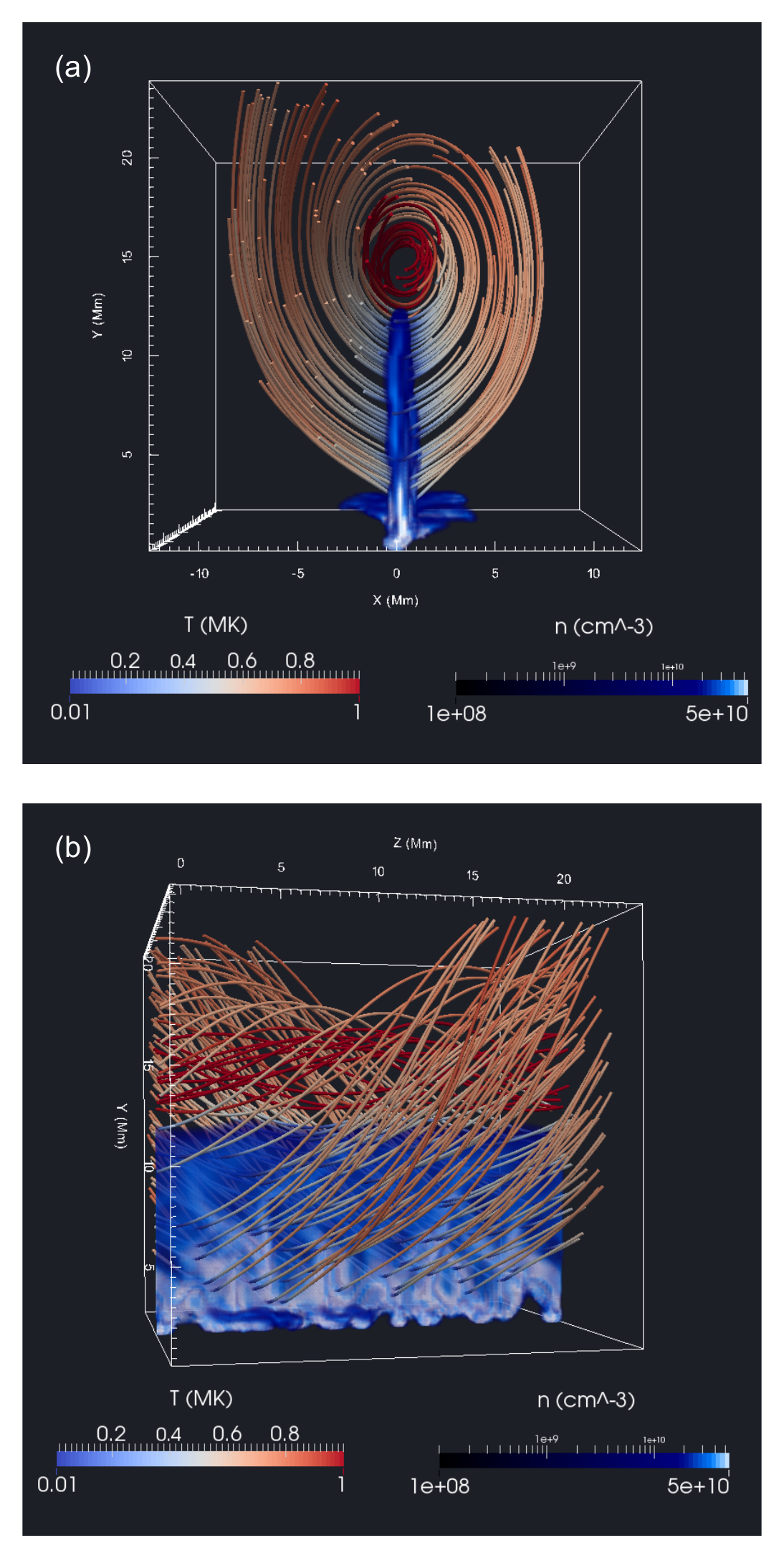}
    \caption{Simulation result at $t=104~\mathrm{min}$ for case R.
      Panels (a) and (b) are snapshots from different angles. The lines represent
      the magnetic field, and the line colors represent the temperature.}
    \label{fig:3dview}
  \end{center}
\end{figure}

\begin{figure}
  \begin{center}
    \includegraphics[bb=0 0 378 555]{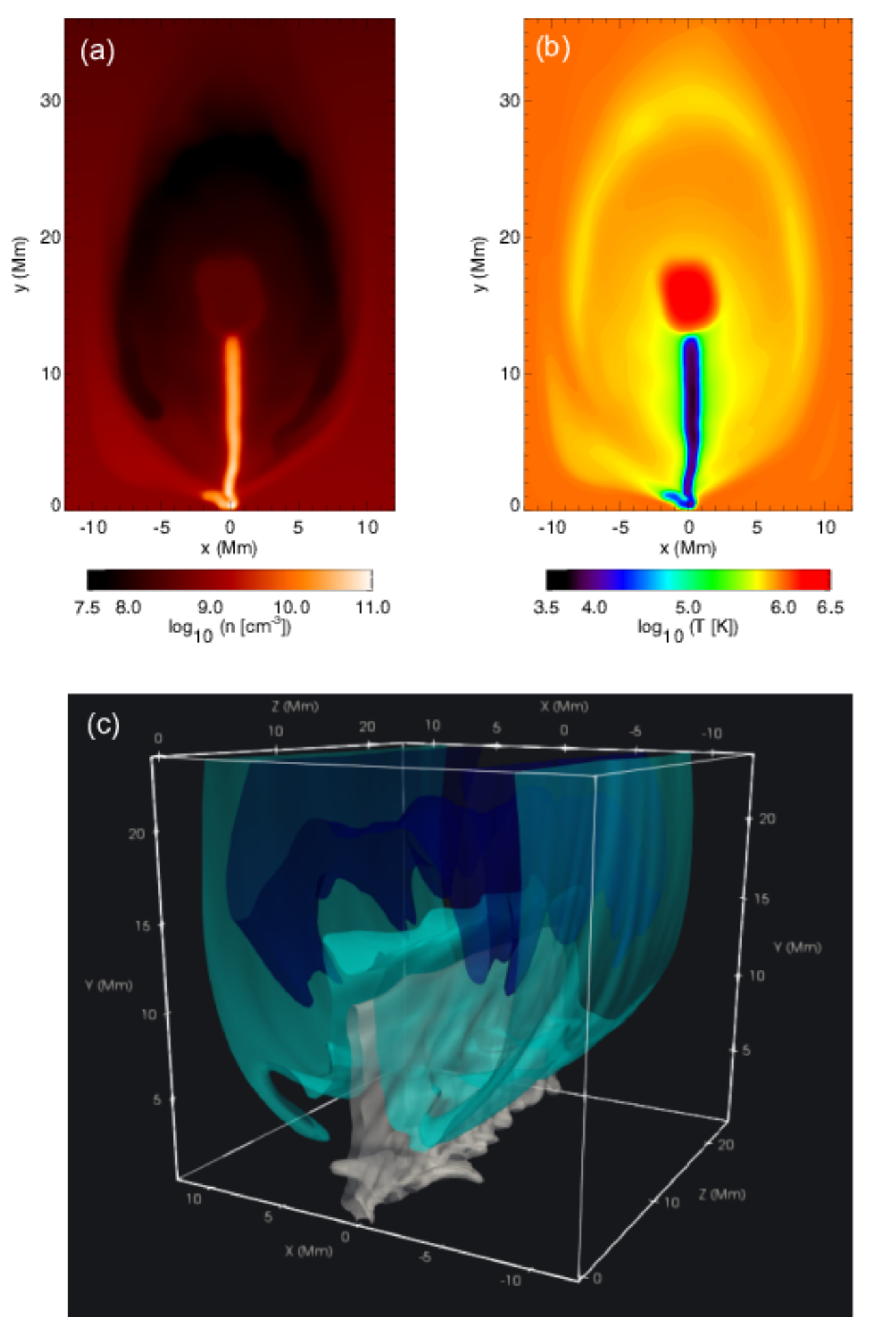}
    \caption{Simulation result at $t=104~\mathrm{min}$ for case R.
      Panels (a) and (b) show the density and temperature
      distributions, respectively, for the $z=12~\mathrm{Mm}$ cross section.
      Panel (c) shows isosurfaces of the density.
      The dark blue, light blue, and white surfaces represent densities of 
      $1.0 \times 10^{8} \mathrm{cm^{-3}}$, $3.5 \times 10^{8} \mathrm{cm^{-3}}$,
      and $1.0 \times 10^{10} \mathrm{cm^{-3}}$, respectively.}
    \label{fig:iso}
  \end{center}
\end{figure}

Figure \ref{fig:3dview} shows three-dimensional snapshots
of the simulation result from different anfig2gles.
The prominence is located along the dips of the flux rope.
The magnetic fields maintain a coherent flux rope structure
even though the local density and velocities evolve in a highly nonuniform manner.
As shown in Fig. \ref{fig:3dview}(b), the vertical threads are not manifestations
of the vertical magnetic fields; rather, they are penetrated by the horizontal magnetic fields.
Figure \ref{fig:iso} shows the density and temperature distribution
in the $z=12~\mathrm{Mm}$ cross section at the same time.
A low-density and high-temperature cavity forms around the prominence
due to the mass depletion after condensation.

\begin{figure}[htbp]
  \begin{center}
    \includegraphics[bb=0 0 452 565]{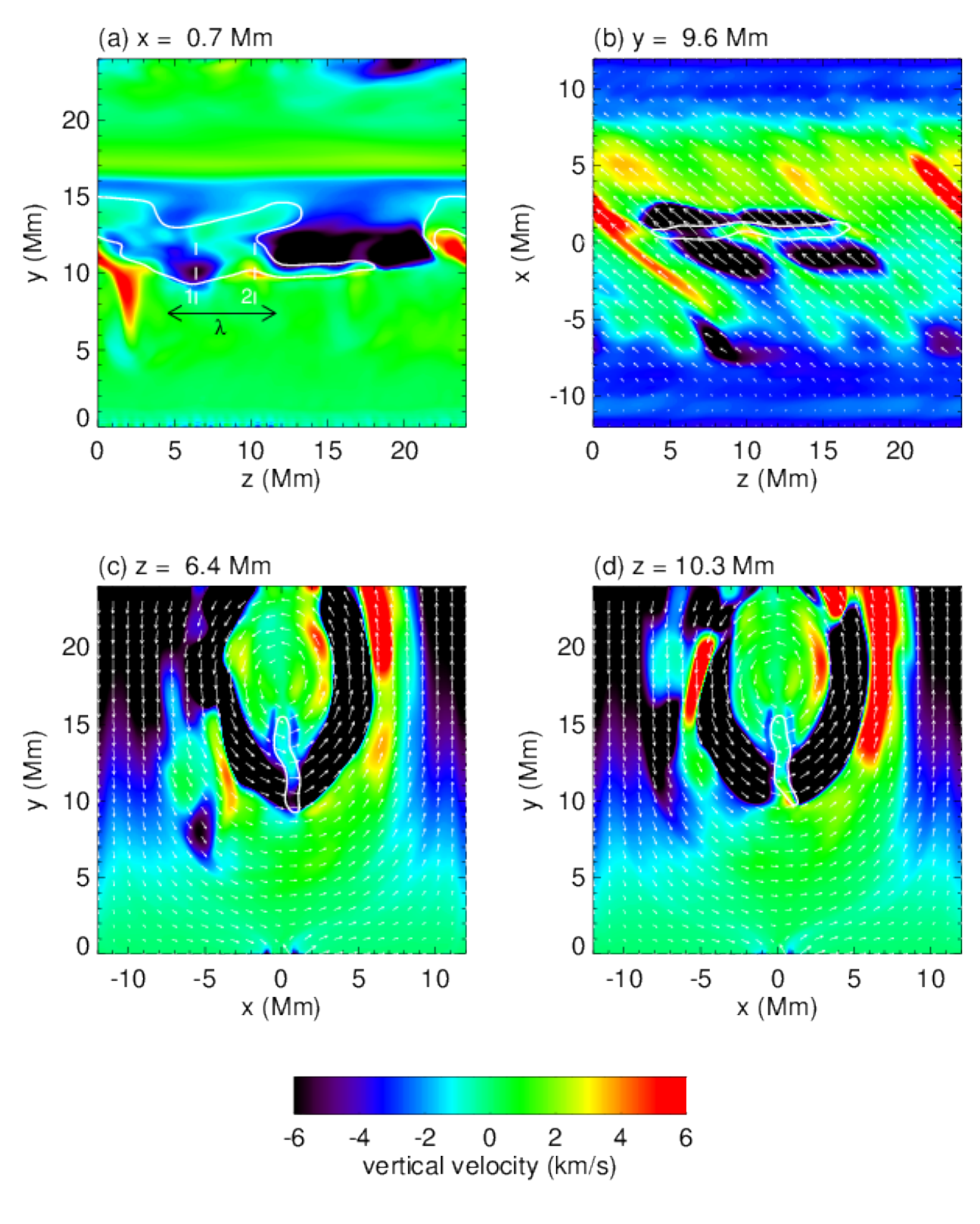}
    \caption{Panels (a), (b), (c) and (d) show snapshots at $t=76~\mathrm{min}$
      in the $x=0.7~\mathrm{Mm}$, $y=9.6~\mathrm{Mm}$, $z=6.4~\mathrm{Mm}$,
      and $z=10.3~\mathrm{Mm}$ planes, respectively.
      The colors represent the vertical velocity $v_{y}$.
      The locations of panels (c) and (d) corresponds to those of dashed lines 1 and 2 in panel (a),
      respectively.
      The solid lines are density isocontour of $n=6 \times 10^{9}~\mathrm{cm^{-3}}$
      indicating the location of prominence. The arrows in panel (b) represent
      the magnetic field vector $(B_{y},B_{z})$, and those in panels (c), and (d)
      represent $(B_{x},B_{y})$.}
    \label{fig:vy}
  \end{center}
\end{figure}

\begin{figure}[htbp]
  \begin{center}
    \includegraphics[bb=0 0 509 184]{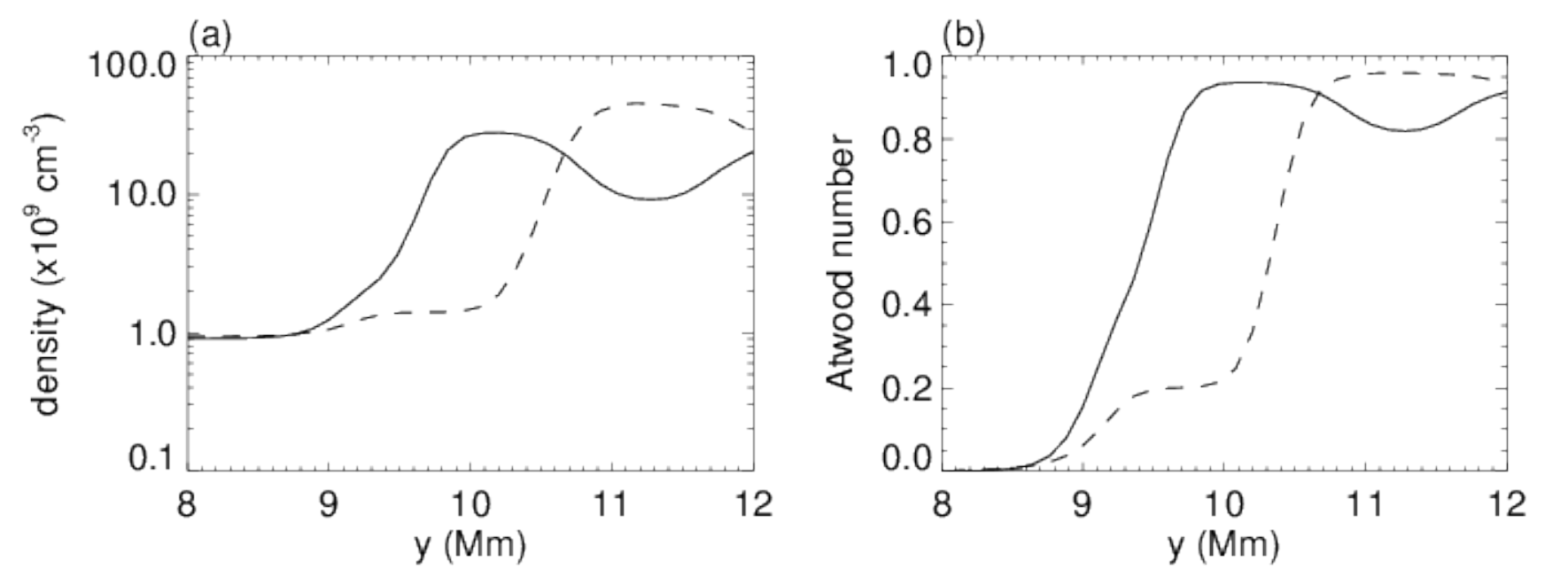}
  \end{center}
  \caption{Profiles of density and the Atwood number
    along the dash lines in Fig. \ref{fig:vy} (a).
    The solid and dashed line represents the values of lines 1 and 2, respectively.}
  \label{fig:atwood}
\end{figure}

\begin{figure}[htbp]
  \begin{center}
    \includegraphics[bb=0 0 509 254]{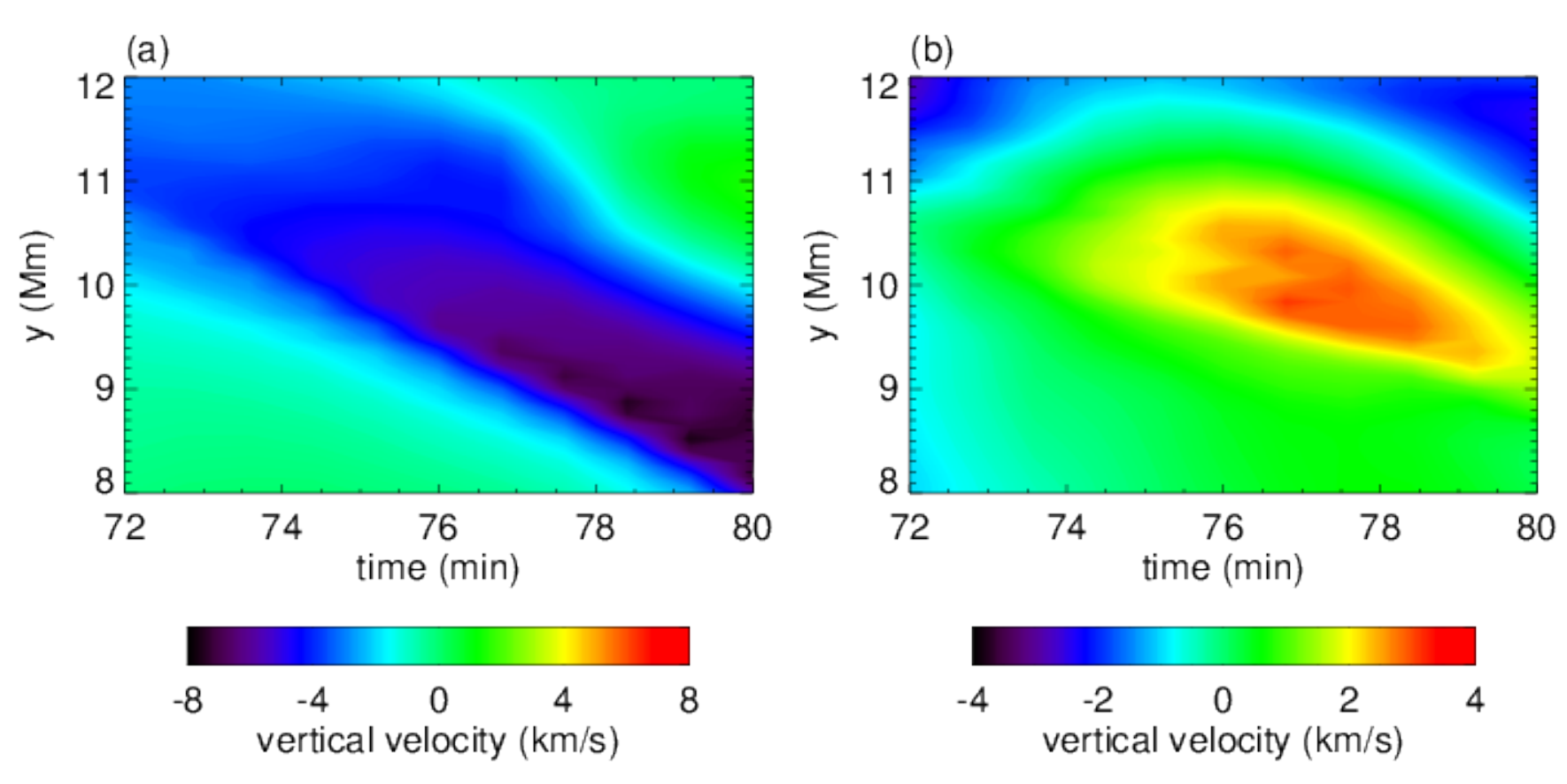}
  \end{center}
  \caption{Panels (a) and (b) show time evolution of the downflow and the upflow along
    the dashed-lines 1 and 2 in Fig. \ref{fig:vy} (a), respectively.
    The colors show the vertical velocity $v_{y}$.}
  \label{fig:vyt}
\end{figure}

\begin{figure}
  \begin{center}
    \begin{tabular}{c}
      \begin{minipage}{\hsize}
        \begin{center}
          \includegraphics[bb=0 0 424 226]{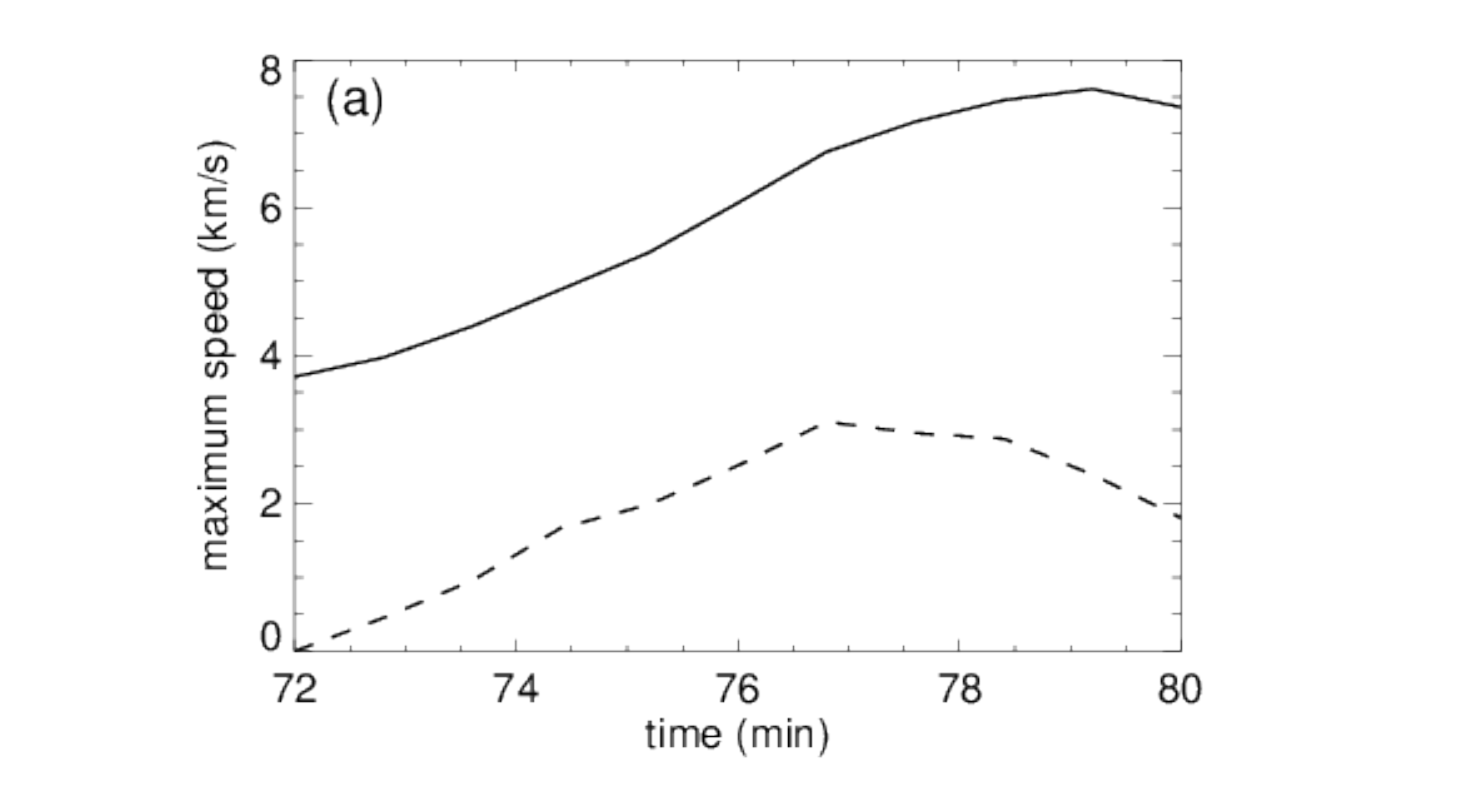}
        \end{center}
      \end{minipage}\\
      \begin{minipage}{\hsize}
        \begin{center}
          \includegraphics[bb=0 0 424 226]{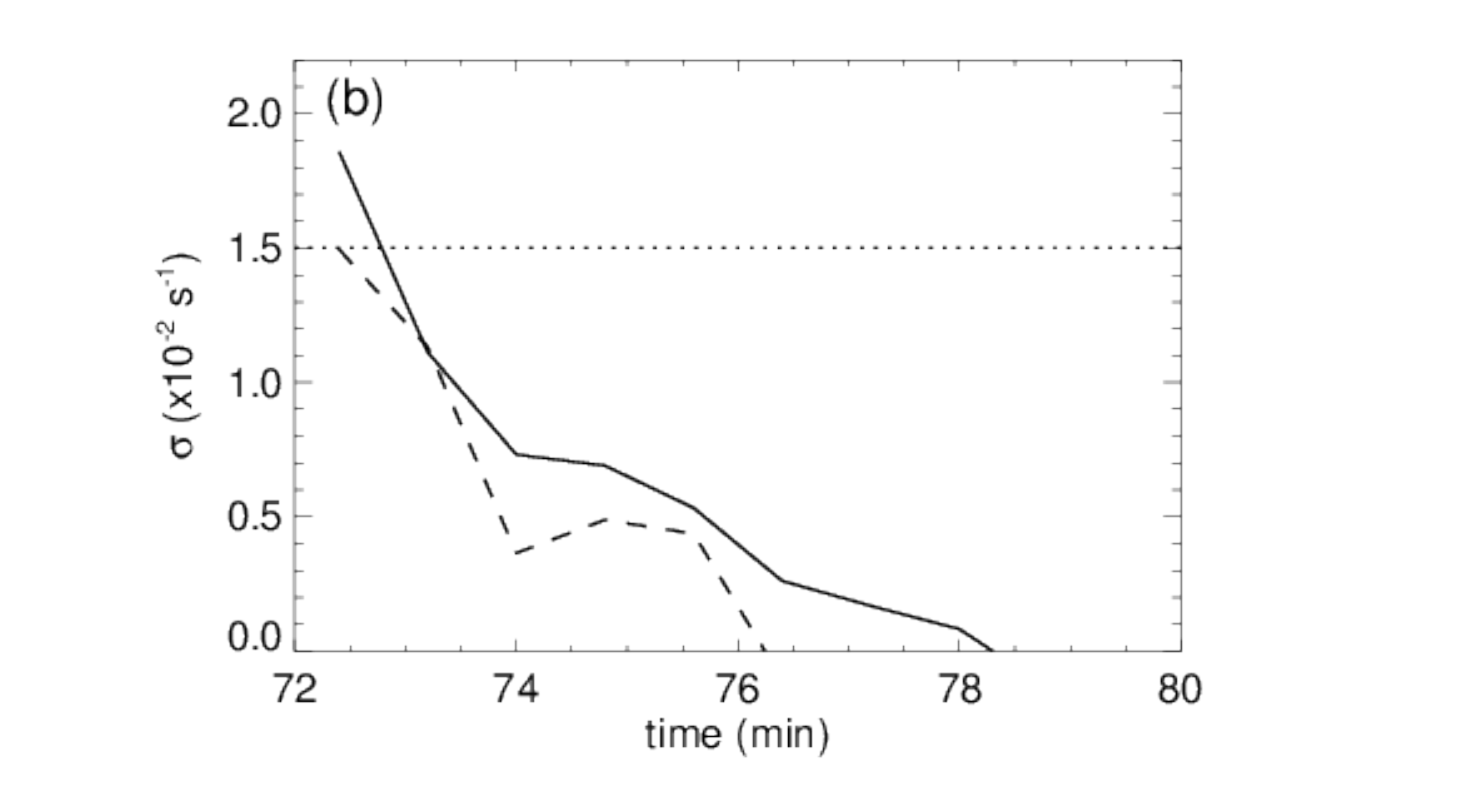}
        \end{center}
      \end{minipage}
    \end{tabular}
    \caption{Panel (a) shows maximum speed of the downflow $V_{\mathrm{down}}$ (solid line) and
      that of the upflow $V_{\mathrm{up}}$ (dashed line) at each time in Fig. \ref{fig:vyt}.
      Panel (b) shows the measured growth rates of the downflow (solid line) and the upflow (dashed line).
      The dotted line represents the analytical growth rate.}
    \label{fig:vyp_sig}
  \end{center}
\end{figure}

Figure \ref{fig:vy} shows the distribution of the vertical flows, location of prominence,
and magnetic field during condensation.
The strong downflows along the coronal magnetic field connected to the prominence
are the condensation flows (panels (b), (c), and (d)). 
Both downflows and upflows are found 
along the interface between the prominence and the lower corona (panels (a) and (d)),
which is indicative of ongoing Rayleigh--Taylor instability.
The growth rates of both flows, which are represented by dashed lines 1 and 2
in Fig. \ref{fig:vy}(a), were compared with the analytical linear growth rate of
the Rayleigh--Taylor instability.
First, the analytical growth rate was estimated. 
The linear growth rate of the magnetic Rayleigh--Taylor instability in a uniform magnetic field
is given by
\begin{equation}
  \sigma ^{2} _{l}=gAk-\frac{\vctr{k}\cdot \vctr{B}}{2\pi \left(\rho _{+}+\rho _{-} \right)},
  \label{eq:lg}
\end{equation}
where $\vctr{k}$ is the wave vector, and $A=(\rho _{+}-\rho _{-})/(\rho _{+}+\rho _{-})$
is the Atwood number, $\rho _{+}$ and $\rho _{-}$ are the densities of heavy and light fluids,
respectively \citep[e.g.][]{Hillier2018RvMPP}.
As shown in Fig. \ref{fig:vy}(b), the wave vectors of the perturbation are virtually perpendicular
to the magnetic field. Hence, we focus on the growth rate of the interchange mode
($\vctr{k}\cdot \vctr{B}=0$).
The wavelength of the perturbation is approximately $\lambda =2\pi /k=7~\mathrm{Mm}$, as shown
by the black arrow representing the interval between the upflows
in Fig. \ref{fig:vy}(a).
Figure \ref{fig:atwood} shows the profile of the density 
and the Atwood number along the dashed lines in Fig. \ref{fig:vy}(a),
where the Atwood number in panel (b) is computed by substituting 
the density at $y=8~\mathrm{Mm}$ with $\rho _{-}$ and the local density with $\rho _{+}$.
The Atwood number of the prominence varies from $0.8$ to $0.95$.
By substituting $A=0.95$ and $\lambda =2\pi /k=7~\mathrm{Mm}$
for the first term of the right side of Eq. (\ref{eq:lg}),
the analytical linear growth rate of the interchange mode
$\sigma _{l}=1.5\times 10^{-2}~\mathrm{s}^{-1}$.
The growth rate of the flows was then measured in the simulation.     
Figure \ref{fig:vyt} depicts the time evolution of both the flows along the dashed lines
in Fig. \ref{fig:vy}(a). At $t=72~\mathrm{min}$, the upflow begins to grow,
and the finite speed of the downflow at $t=72~\mathrm{min}$ is due to the condensation flows.   
Figure \ref{fig:vyp_sig}(a) depicts the time evolution of the maximum speed of
the downflow $V_\mathrm{down}$ and that of the upflow $V_\mathrm{up}$ in Fig. \ref{fig:vyt}.
The growth rate $\sigma $ of both the flows can be estimated by
\begin{equation}
  \sigma = \frac{1}{V}\frac{dV}{dt},
\end{equation}
where $V=V_\mathrm{up}$ is substituted for the upflow, and  
$V=V_\mathrm{down}-V_{c}$ is substituted for the downflow, where $V_{c}$ is the speed
of the condensation flow.
We assume that $V_{c}=V_\mathrm{down}(t=72~\mathrm{min})$ does not change
in the short duration.
Figure \ref{fig:vyp_sig}(b) shows the measured growth rates of both the flows,
which are comparable with the analytical growth rate.
This suggests that a mechanism similar to the Rayleigh--Taylor instability
(RTI-like mechanism) facilitates the corrugation of the interface
between the prominence and the lower corona.
The situation reproduced
in our simulation is not the rigorous form of the magnetic Rayleigh--Taylor instability
discussed in the previous studies because the condensation flows coexist in the interface,
indicating that the system is not in a state of complete thermal and mechanical equilibrium.
As shown in Fig. \ref{fig:vy}(d), the upflows are located between the condensation downflows
(see around $(x,y)=(1~\mathrm{Mm},10~\mathrm{Mm})$), and are gradually canceled.
Therefore, the upflows do not evolve into the dark plumes observed
in prominences \citep{Berger2008ApJ,Berger2010ApJ}.
Note that the acceleration of the downflow in our simulation
is estimated to be $-10~\mathrm{m~s^{-2}}$ 
from the inclination of $V_\mathrm{down}$ in Fig. \ref{fig:vyp_sig}(a),
which is significantly smaller than the gravitational acceleration $g=-270~\mathrm{m~s^{-2}}$.

\begin{figure}
  \begin{center}
    \includegraphics[bb=0 0 452 496]{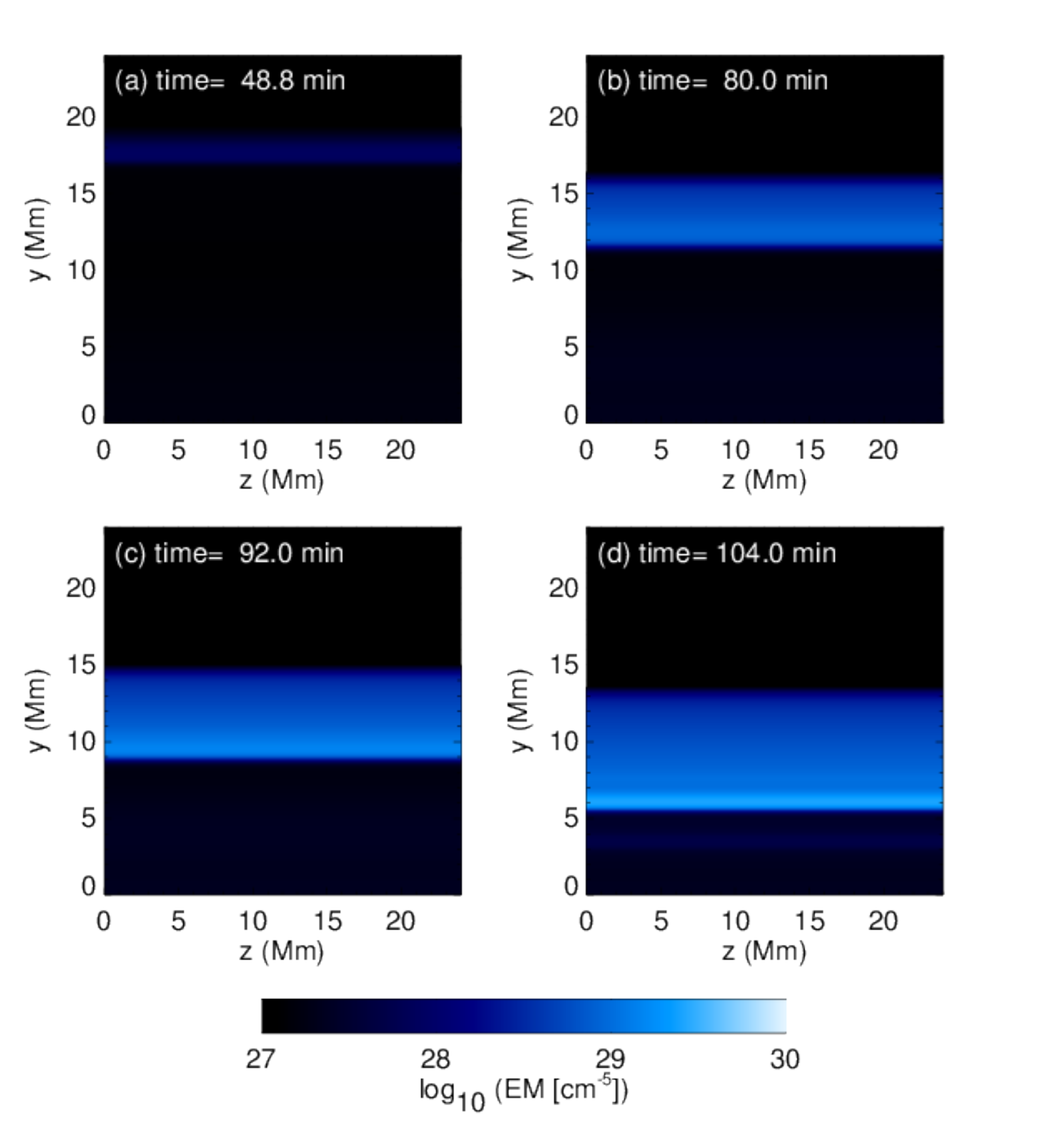}
    \caption{Time evolution of the emission measure along the $x$-axis in case N.}
    \label{fig:2dview}
  \end{center}
\end{figure}

\begin{figure}
  \begin{center}
    \begin{tabular}{c}
      \begin{minipage}{\hsize}
        \begin{center}
          \includegraphics[bb=0 0 424 226]{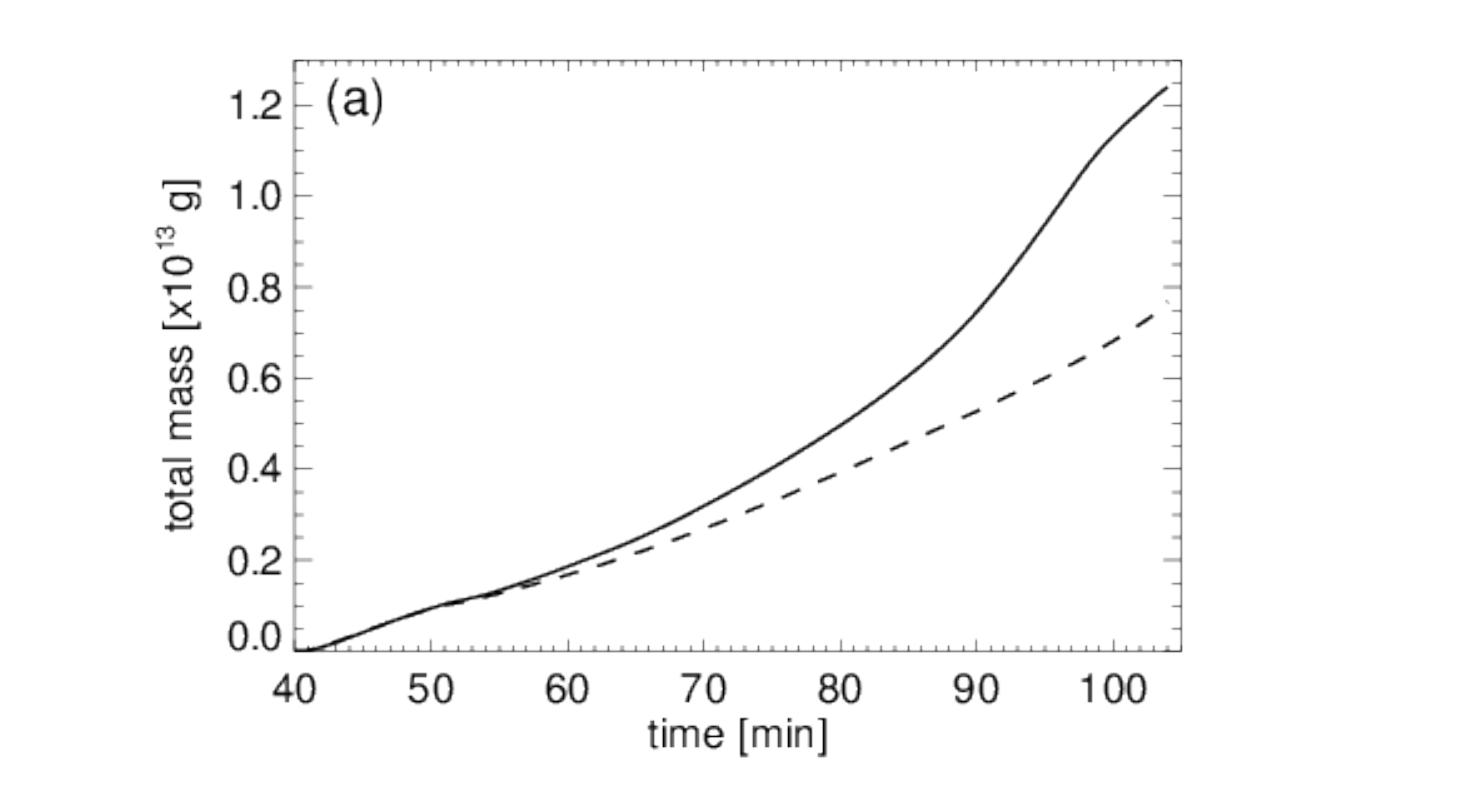}
        \end{center}
      \end{minipage}\\
      \begin{minipage}{\hsize}
        \begin{center}
          \includegraphics[bb=0 0 424 226]{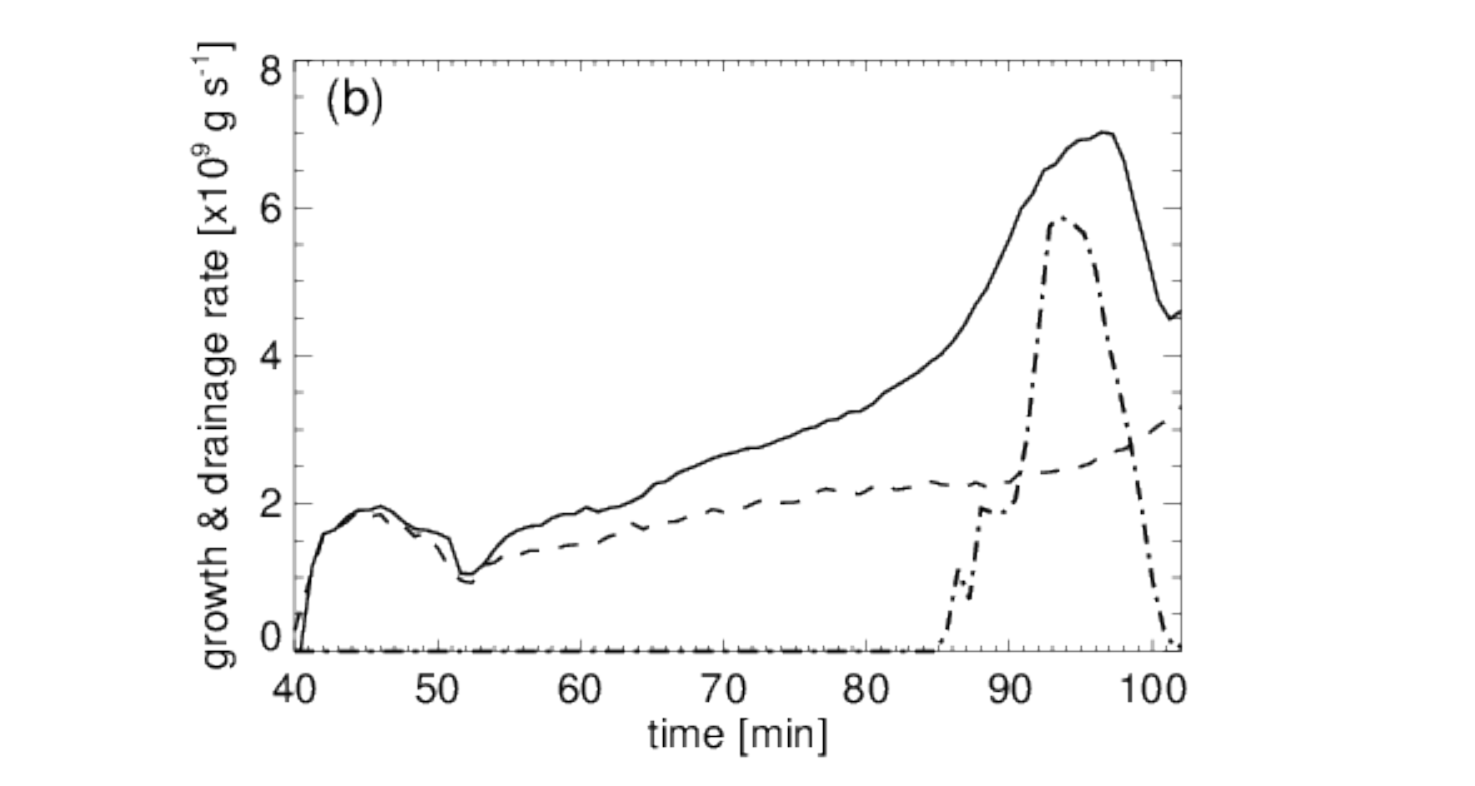}
        \end{center}
      \end{minipage}
    \end{tabular}
    \caption{Panel (a) shows the temporal evolution of the total mass
      of cool plasma $(T<10^{5}~\mathrm{K})$.
      The solid and dashed lines represent cases R and N, respectively.      
      Panel (b) shows the mass growth rate and the drainage rate.
      The solid and dashed lines represent the growth rates in case R and N, respectively.
      The dash-dotted line represents the mass drainage rate in case R.}
    \label{fig:mass_tevol}
  \end{center}
\end{figure}

To investigate the impact of the dynamic state on the condensation rate,
we compare the cases with and without dynamic state.
Figure \ref{fig:2dview} shows snapshots of the simulation results for case N
in which the footpoint motion does not contain random components.
Due to the constraints of the 2.5-dimensional assumption, all variables are uniform
in the $z$-direction.
The temporal evolution of the total mass and the mass growth rate of the prominences
in cases R and N are shown in Fig. \ref{fig:mass_tevol}.
The mass growth rate of the prominence is computed as
\begin{equation}
  \dot{M}_{\mathrm{prom}}=\frac{d}{dt}\int_{T<10^{5}\mathrm{K}} \rho (x,y,z) dx dy dz.
\end{equation}
Until $t=50~\mathrm{min}$, the mass growth rates are the same for these two cases.
After $t=50~\mathrm{min}$, the mass growth rate in case R becomes larger than that in
case N. As shown in Fig. \ref{fig:mass_tevol}(b), the mass growth rate in case R
is enhanced around $t=80~\mathrm{min}$ just after the spikes begin to evolve
(see also Fig. \ref{fig:ems3d}(b)).
The mass drainage rate is also enhanced with the evolution of the spikes.
The dash-dotted line in Fig. \ref{fig:mass_tevol}(b) shows the temporal evolution
of the mass drainage rate computed as
\begin{equation}
  \dot{M}_{\mathrm{drain}}=\int _{T<10^{5}\mathrm{K}} \rho v_{y}(x,y_{c},z)dxdz
\end{equation}
where $y_{c}=5~\mathrm{Mm}$ is selected.
The mass drainage rate is comparable to and temporally correlated with the condensation rate.
Our simulation results support the observational findings of \citet{Liu2012ApJ}
and give insights into the self-induced mass maintenance mechanism via the coupling of
the radiative condensation and the Rayleigh--Taylor instability.

\begin{figure}
  \begin{center}
    \includegraphics[bb=0 0 339 226]{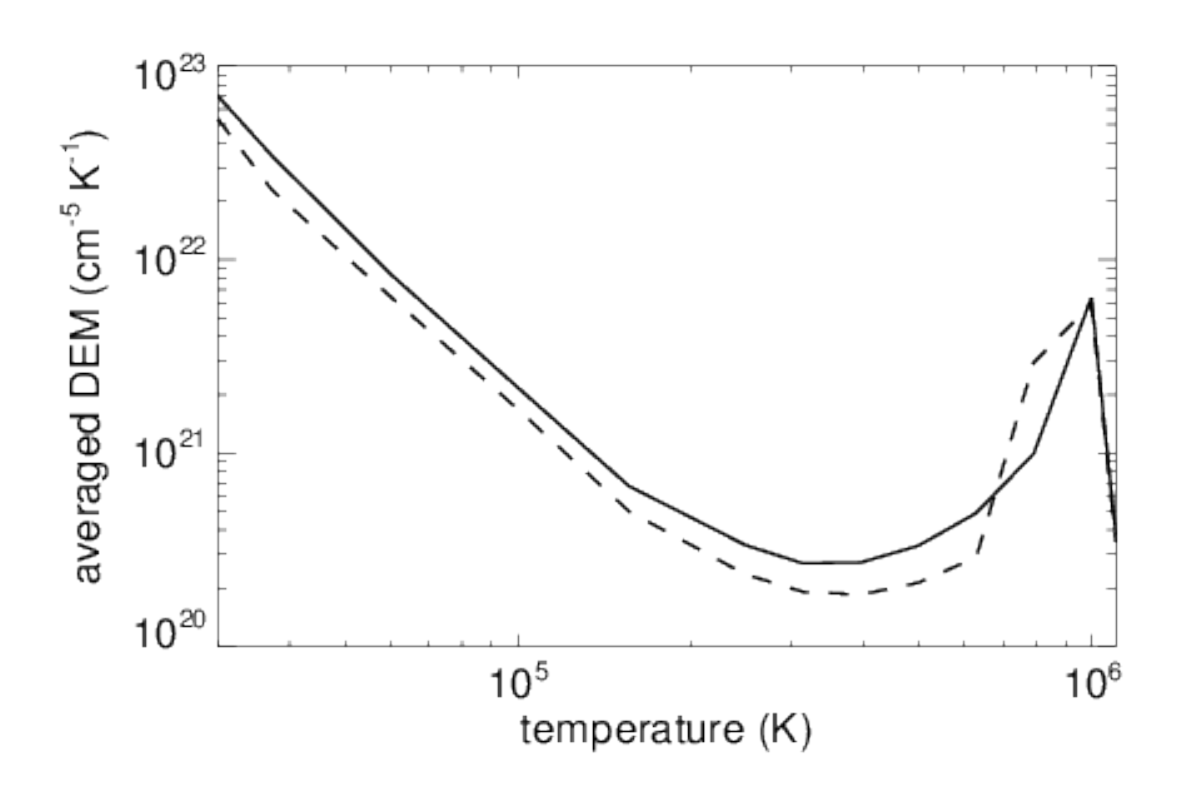}
    \caption{Averaged DEMs during condensation.
      The solid and dashed lines represent the DEMs of case R at $t=88.0~\mathrm{min}$
      and that of case N at $t=97.6~\mathrm{min}$, respectively.}
    \label{fig:dems}
  \end{center}
\end{figure}

\begin{figure}
  \begin{center}
    \includegraphics[bb=0 0 509 254]{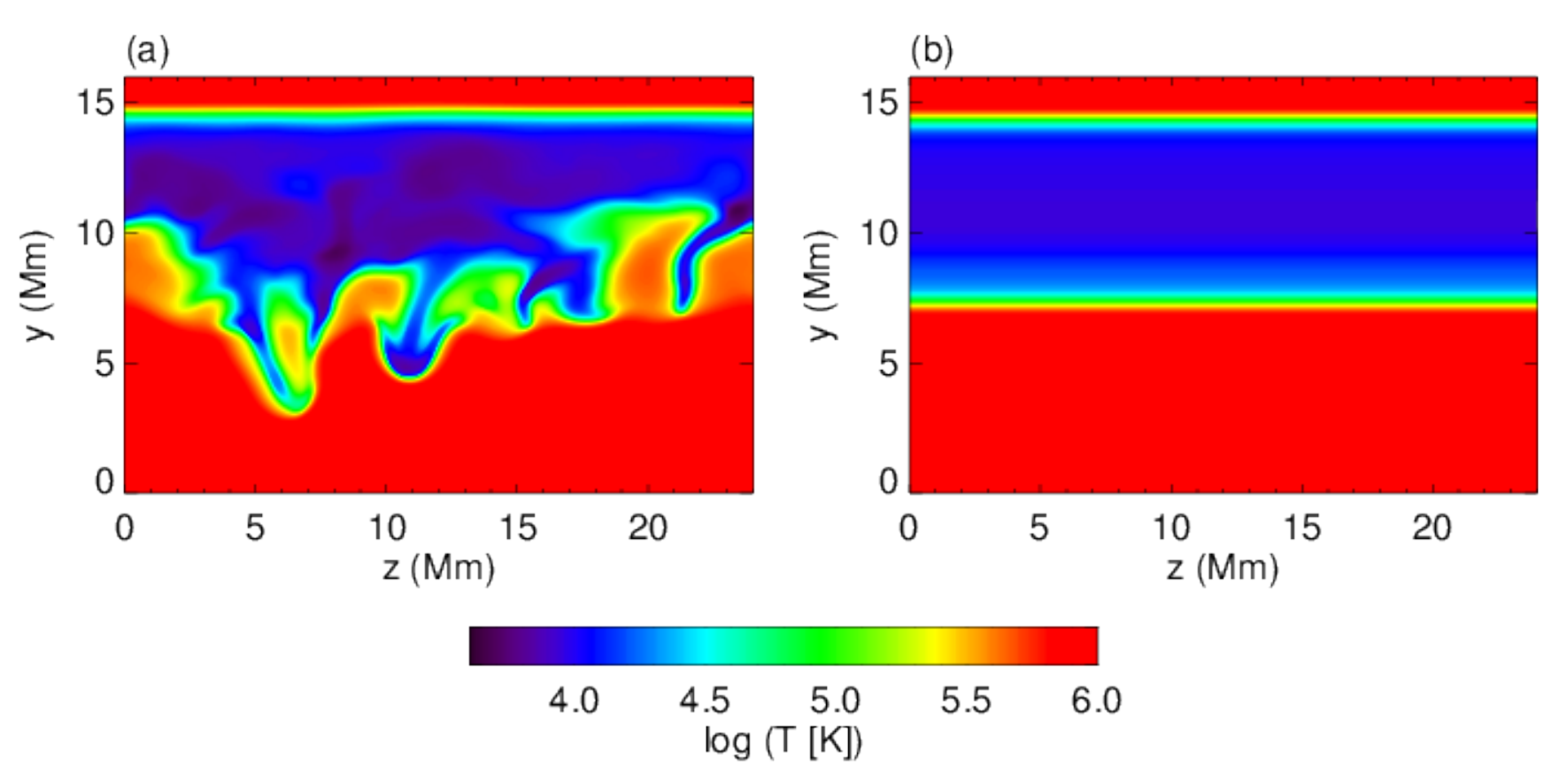}
    \caption{Snapshots of simulation results in the $x=0.3~\mathrm{Mm}$ plane.
      The colors represent the temperature.
      Panels (a) and (b) show the snapshots of case R at $t=88.0~\mathrm{min}$
      and that of case N at $t=97.6~\mathrm{min}$, respectively.}
    \label{fig:te}
  \end{center}
\end{figure}

Figure \ref{fig:dems} shows the differential emission measures (DEMs)
during condensation for cases R and N defined as
\begin{equation}
  \mathrm{DEM}(T)=n^{2}\frac{dx}{dT},
\end{equation}
where the line of site is in the $x$-direction.
The DEMs are averaged over $0 < y < 14~\mathrm{Mm}$ and $0 < z < 24~\mathrm{Mm}$.
We compared the DEM in case R at $t=88.0~\mathrm{min}$ to that 
in case N at $t=97.6~\mathrm{min}$ because
the total mass of the plasmas cooler than $0.8~\mathrm{MK}$ are the same for those times.
In case R, the DEM in the lower temperature region ($0.1~\mathrm{MK}<T< 0.65~\mathrm{MK}$)
is larger than that in case N.
Figure \ref{fig:te} shows the temperature in the $x=0.3~\mathrm{Mm}$ plane in each case.
In case R, the prominence--corona transition region (PCTR) is deformed
by the Rayleigh--Taylor instability and the area of
the lower temperature region
(the area represented by the green to yellow colors in Fig. \ref{fig:te})
is broader than that in case N.
The difference in the DEM profiles is therefore due to
the deformation via the Rayleigh--Taylor instability in the nonlinear regime.
The radiative cooling rate of the plasmas in the temperature range of
$0.3~\mathrm{MK} < T < 0.5 ~\mathrm{MK}$ is higher than those for other temperatures.
Because a larger mass is distributed in this temperature region,
condensation proceeds more efficiently in case R than in case N.
    
\section{Discussion} \label{sec:discussion}

We reproduced a prominence with vertical threads and internal flows  
within the framework of the reconnection--condensation model.
In the model,
reconnection via footpoint conversing motion creates a flux rope structure,
and the radiative condensation inside the flux rope leads to formation of prominence.
In contrast to the homogeneous footpoint motion in
the previous simulations \citep{KanekoYokoyama2015ApJ,KanekoYokoyama2017ApJ},
the speed of the footpoint motion varies randomly along the PIL in the present simulation.
Because of the density and velocity fluctuations inside the flux rope
given by the random footpoint speed, radiative condensation heterogeneously proceeds.
As a result, the prominence with highly dynamic interior was reproduced. 
We confirmed that the RTI-like mechanism facilitates the corrugation of the interface
between the prominence and the lower corona.
The descending spikes reflected at the bottom boundary generates the upflows.
Due to the collision of the downflows and  the upflows, the spikes evolves into the thin vertical threads.
The downward speeds of the spikes and the widths of the vertical threads are consistent
with the typical values in observations.

We found that the mass condensation rate is enhanced in a dynamic state.
The condensation rate was comparable to the mass drainage rate
of the downflows in our simulation.
Significant mass drainage via downflows has actually
been found in observations of prominences \citep{Zirker1994SoPh,Liu2012ApJ}.
A study by \citet{Liu2012ApJ} suggested
that the condensation rate is comparable to the mass drainage rate
to maintain the total mass of the prominence.
Our results support this suggestion from the observations.
In a recent simulation by \citet{XiaKeppens2016ApJ}, the mass circulation
between the chromosphere and the corona via a prominence was reproduced
based on the evaporation--condensation model.
However, the presence of the Rayleigh--Taylor instability was unclear
in their fragmented condensations.
It is also likely that the condensation rate in their simulation could be enhanced
because the fragmentation extends the total volume of PCTRs.

In our simulation, the thin vertical threads form after the spikes are reflected
at the bottom boundary, where the velocities are numerically fixed to zero.
Previous MHD simulation including the chromosphere showed that
the materials in falling spikes of the Rayleigh--Taylor instability
are reflected at the top of the chromosphere and returned to coronal heights \citep{KeppensXia2015ApJ}.
It is possible to assume the artificial boundary in our simulation as the interface
between the chromosphere and the corona.
In observations, the complicated flow pattern in the prominences
suggests that the collisions between internal flows are likely to occur;
however, the reflection of downflows at the chromosphere has not been clearly detected.

The observed dark plumes \citep{Berger2008ApJ,Berger2010ApJ}
are not reproduced in our simulation.
In observations \citep{Berger2011Natur,Dudik2012ApJ,Berger2017ApJ},
the dark plumes (corresponding to the bubbles of the Rayleigh--Taylor instability)
originated at the interface between the prominence and the buoyant cavity,
which is a semicircular dark region between the prominence and the chromosphere.
The buoyant cavities (also called prominence bubbles)
were conjectured to be emerging fluxes.
To explain the origin of dark plumes inside prominence bodies,
it might be necessary to consider additional emerging fluxes interacting with the flux ropes.

In observations,
a possible origin of the footpoint converging motion
is the collisions of supergranular diverging flows, 
which have typical speeds of $0.3~\mathrm{km/s}$ on the photosphere
and lifetimes of one day \citep{Rondi2007AA,Schmieder2014AA}.
In the present simulations, we set the footpoint motion with a speed of
$v_{00}=12~\mathrm{km/s}$ and a duration of $1440~\mathrm{s}$.
The footpoint speed in our simulations is much faster than the typical values
in the observations; however, the migration distance of the magnetic components
(the product of the speed and duration) is comparable.
The footpoint speed in our simulation is still much slower
than the sound speed and the Alfv\'{e}n speed in the corona.
It is likely that this result will not change significantly
even if the typical observational speed is applied.

In our simulation, the descending spikes (and the perturbation at the interface)
have a spatial scale of approximately $7~\mathrm{Mm}$.
It was difficult to clarify the origin of this spatial scale from a single simulation.
However, it is possible that the spatial scale is the wavelength of the interchange mode
of the Rayleigh--Taylor instability, which depends on the numerical viscosity in the simulation.
To estimate the effect of numerical viscosity,
the simulation results must be compared at different resolutions.
It is also important to note that
the growth of the upflow bubbles and the descending spikes may be
asymmetrically influenced by radiative condensation,
whereby the condensation flows cancel out the upflows and amplify the downflows.
To clarify the effect, a simpler model including
an equilibrium interface is needed, and a parameter survey of the perturbation wavelength
should be performed.

The development of MHD simulations that includes a self-consistent ionization effect
is one of the important future steps for the realistic numerical modeling. 
Numerical study of the magnetic Rayleigh--Taylor instability including ambipolar diffusion
have indicated that the growth rate and the flow speed are affected by the cross-field diffusion
\citep{Khomenko2014AA}.
The analytical studies by \citet{Low2012ApJa,Low2012ApJb} also suggest that
prominence condensation inevitably creates a discrete current, leading to the spontaneous cross-field
mass transport with the presence of neutrals.
The ionization effect must be considered to obtain a more accurate evaluation of mass flux
associated with downflows.

Recent observational studies have revealed the turbulent properties  
in quiescent prominences \citep{Leonardis2012ApJ,Freed2016ApJ,Hillier2017AA}.
A break in the scaling exponent has been found to exist
at a spatial scale of $2000~\mathrm{km}$ in the power spectrum or structure function.
Our simulation was unable to reproduce such a turbulent nature
probably due to the numerical viscosity. Simulations with higher resolution
are required to understand the turbulent characteristics of solar prominences.

\section{Conclusions}\label{sec:conclusion}
We reproduced a prominence with vertical threads and internal downflows
within the framework of the reconnection--condensation model.
In the present model, footpoint motion with a random speed along a PIL is given.
The Rayleigh--Taylor instability is triggered during radiative condensation.
The spikes eventually evolve into descending pillars and thin vertical threads.
It was found that the mass condensation rate is enhanced 
to the same level as the mass drainage rate in dynamic state.
The extension of the PCTR at lower temperatures, where the radiative loss is the highest,
leads to higher mass condensation rate.
Our results reveal the impact of the dynamic state on the radiative condensation rate
and support the observational hypothesis claiming
a balance between the condensation rate and the mass drainage rate in prominences.

\acknowledgments

This work was supported by MEXT/JSPS KAKENHI Grant Numbers JP16J06780, JP15H03640
and 15H05814.
Numerical computations were conducted on a Cray XC30 supercomputer at the
Center for Computational Astrophysics (CfCA) of the National Astronomical
Observatory of Japan.


\end{document}